\documentclass[preprint,12pt]{elsarticle}

\usepackage{graphicx}

\usepackage{amssymb,amsmath,dsfont}

\newcounter{bla}

\journal{Computer Physics Communications}

\usepackage{enumitem}
\usepackage{url}
\usepackage[export]{adjustbox}
\newcommand{\eg}{\textit{e.g.\ }}
\newcommand{\etal}{\textit{et al.\ }}
\newcommand{\ie}{\textit{i.e.\ }} 

\newcommand{\rmi}{\mathrm{i}} 
\newcommand{\rme}{\mathrm{e}}
\newcommand{\rmd}{\mathrm{d}}

\newcommand{\bra}[1]{\left<#1\right|}
\newcommand{\ket}[1]{\left|#1\right>}
\newcommand{\braket}[2]{\left<#1|#2\right>}

\begin{document}

\begin{frontmatter}



\title{QSWalk: a {\it Mathematica} package for quantum stochastic walks on
  arbitrary graphs}


\author{Peter E. Falloon}
\author{Jeremy Rodriguez}
\author{Jingbo B. Wang\corref{author}}

\cortext[author] {Corresponding author.\\\textit{E-mail address:} jingbo.wang@uwa.edu.au}
\address{School of Physics, The University of Western Australia, Crawley WA 6009, Australia}

\begin{abstract}
  We present a {\it Mathematica} package, {\tt QSWalk}, to simulate the time evaluation of
  Quantum Stochastic Walks (QSWs) on arbitrary directed and weighted
  graphs. QSWs are a generalization of continuous time quantum walks that
  incorporate both coherent and incoherent dynamics and as such, include both
  quantum walks and classical random walks as special cases. The incoherent
  component allows for quantum walks along directed graph edges. The
  dynamics of QSWs are expressed using the Lindblad formalism,
  originally developed for open quantum systems, which frames the problem in the
  language of density matrices. For a QSW on a graph of $N$ vertices, we have a sparse superoperator in 
  an $N^2$-dimensional space, which can be solved
  efficiently using the built-in {\tt MatrixExp} function in {\it
    Mathematica}. We illustrate the use of the {\tt QSWalk} package through
  several example case studies.
\end{abstract}


\end{frontmatter}



{\bf PROGRAM SUMMARY}

\begin{small}
\noindent
{\em Program Title:} QSWalk.m                           \\
{\em Licensing provisions:} none                                 \\
{\em Programming language:} {\it Mathematica}. \\
{\em Computer and operating system:} any system running {\it Mathematica}
Version 10 (or later).  \\
{\em RAM:} depends on graph size.                                 \\
{\em Classification:} 4.15.                                        \\
{\em Nature of problem:}\\
  The {\tt QSWalk} package provides a method for simulating quantum stochastic walks on
  arbitrary (directed/undirected, weighted/unweighted) graphs.
   \\
{\em Solution method:}\\
  For an $N$-vertex graph, the solution of a quantum stochastic walk can be
  expressed as an $N^2\times N^2$ sparse matrix exponential. The {\tt QSWalk} package makes
  use of {\it Mathematica}'s sparse linear algebra routines to solve this
  efficiently.
   \\
{\em Restrictions:}\\
  The size of graphs that can be treated is constrained by available memory.
   \\
{\em Running time:}\\
  Running time depends on the size of the graph and the propagation time of the
  walk.
   \\
\end{small}

\section{Introduction}

Over the past two decades, quantum computing and information theory have
emerged as one of the most exciting and promising frontiers of research, at the
forefront of both quantum physics and computer science (see, \eg
\cite{nielsen2010} and references therein). One particularly interesting avenue
of research is the {\it quantum walk} (QW) \cite{kempe2003, kendon2007,
  venegasAndraca2012, manouchehri2014}, a quantum
mechanical analogue of the classical random walk (CRW). CRWs model the random motion of a ``walker'' over the vertices of some
graph, with each step chosen at random from the edges connecting to adjacent
vertices \cite{rudnick2004, blanchard2011}. The CRW is ubiquitous throughout
physics and applied mathematics, being associated with phenomena as diverse as
Brownian motion in liquids \cite{einstein1956} and price fluctuations in
financial markets \cite{malkiel1999}.

QWs were first described by Aharanov in 1993 \etal \cite{aharonov1993}, who
introduced a discrete-time formulation in which random steps are effected by
means of a unitary {\it coin operator} \cite{aharonov2001, ambainis2001}. A
continuous-time model, with no need of a coin operator, was subsequently
introduced by Farhi and Gutmann in 1998 \cite{farhi1998}.  The key difference of
both discrete- and continuous-time QWs with respect to the equivalent CRW is that
the QW evolves according to quantum amplitudes instead of probabilities, and so
the diffusion process of the CRW is replaced by a phase-coherent unitary
evolution in which different paths can interfere. This phase-coherence leads to
drastically different behaviour, which has been exploited to develop algorithms
on QWs achieving an exponential speedup {\it vis-{\`a}-vis} the corresponding
CRW \cite{childs2003}.

A significant limitation of QWs is that the time evolution must be unitary in
order to conserve probability. This presents a particular challenge when
attempting to apply the QW to directed graphs, \ie graphs containing edges which
can only be traversed in one direction. While there have been several approaches
to extend discrete-time QWs to directed graphs \cite{szegedy2004,
  montanaro2007}, these do not carry over to the continuous-time case.

One approach that does yield a version of continuous-time QWs applicable to
directed graphs is the so-called Quantum Stochastic Walk (QSW), a generalization
of both QWs and CRWs developed by Whitfield \etal \cite{whitfield2010}. The
object of study in a QSW is the density matrix, which combines the coherent
dynamics of quantum mechanical states with the incoherent evolution of a
probability vector. The QSW posits a master equation comprising a combination of
the QW Hamiltonian, representing unitary evolution, and a term describing
incoherent scattering. The latter is based on the Lindblad formalism from open
quantum systems theory \cite{breuer2002, rivas2011}. By tuning the relative
weighting of these components, the QSW can describe a continuum of behavior with
the QW and CRW at either end. Importantly, the incoherent scattering component
of the QSW is not subject to the same constraints as the Hamiltonian of the QW
and is thus able to represent irreversible scattering, \eg along one-way edges
in directed graphs. QSWs have recently been applied to several problems at the
forefront of current quantum information theory research, including the quantum
PageRank algorithm \cite{sanchezBurillo2012}, quantum neural networks
\cite{schuld2014} and, most recently, decision-making networks
\cite{martinezMartinez2016}.

In this paper we present a {\it Mathematica} package, {\tt QSWalk}, for the
evaluation of QSWs on arbitrary directed graphs. The numerical approach used in
our package is based on vectorization of the density matrix and master equation, allowing a solution in terms of a matrix exponential for which efficient
sparse matrix methods can be used. By writing our package in {\it Mathematica}
we are able to make use of its extensive built-in functionality for graphs and
linear algebra, making our implementation straightforward and
efficient. Furthermore, {\it Mathematica} offers the advantage of an intuitive
notebook environment and visualization functions, and is thus an ideal tool for
exploring QSWs for either research or educational purposes.

Although a python library for quantum walks has recently been made available
\cite{izaac2015}, to our knowledge there is currently no software in the public
domain for the computation of QSWs. A python package, QuTiP, has been written
for the study of open quantum systems \cite{johansson2012, johansson2013}, which
could potentially provide similar functionality to our package. However, it is
not specifically tailored to the context of quantum walks on graphs, and it
requires working in Python, so may not be as accessible as our {\it Mathematica} package.

The remainder of this paper is structured as follows. In the next section we
give an overview of the basic theory underlying QSWs. In Section
\ref{section:package} we describe the {\tt QSWalk} package and some of the
computational methods used therein. We then present several illustrative
applications, including a study of FMO complex in photosynthesis and the quantum Page Rank algorithm, in Section
\ref{section:applications}.

\section{Theory}
\label{section:theory}

Our goal in this section is to present the background theory relevant to QSWs in a
concise and self-contained way. We also aim to establish a clear and consistent
notation, which is particularly important owing to the many different (and sometimes
mutually contradictory) notations in the literature (ours is most similar to that of
Refs.~\cite{whitfield2010} and \cite{childs2002}). We start by briefly defining
graphs and the associated CRWs, followed by QWs, and finish with a more detailed
description of QSWs. For the sake of brevity, and because they are irrelevant to
QSWs, we omit the discrete-time CRW and QW models and discuss only the
continuous-time case. Throughout this section, and the rest of the paper, we
work in units with $\hbar=1$.

\subsection{Classical random walks on graphs}
\label{subsection:CRW}

For our purposes, a graph $G$ is defined as a set of vertices (or {\it nodes})
$\{1,\ldots,N\}$ together with an $N\times N$-dimensional {\it adjacency matrix}
$A$ describing the connections (or {\it edges}) between them.  In the simplest
case of an {\it unweighted} graph, $A_{ij}=1$ if there is an edge from $j$ to
$i$ and $0$ otherwise.  For a $weighted$ graph, on the other hand, edges have
arbitrary weights $A_{ij}>0$ representing the strength of the connection from
$j$ to $i$.  If $A$ is symmetric, so that $A_{ij}=A_{ji}$ for all $i,j$, $G$ is
said to be {\it undirected}; otherwise it is a {\it directed} graph (or
digraph).  For each vertex, the sum of weights for all edges leaving it is
called its {\it out-degree}: $\mathrm{outDeg}(j) = \sum_{i\neq j}A_{ij}$.

We note that the diagonal entries $A_{ii}$ represent {\it self-loops}, edges
which start and finish at the same vertex; if $G$ has no self-loops it is called {\it
  simple}. In the literature on QWs and QSWs simple graphs are usually
considered, however, non-simple graphs can in principle be used.

A (continuous-time) CRW on $G$ may be defined as an $N$-component probability
vector $\mathbf{p}(t)$ whose components $p_i(t)$ sum to unity and represent
the probability for the hypothetical walker being found at vertex $i$ at time
$t$. The initial condition typically places the walker on a particular vertex at
time zero, $p_i(0) = \delta_{i,q}$ for some $q$. The time evolution of
$\mathbf{p}(t)$ is governed by a {\it master equation}:
\begin{equation}\label{eq:CRW master}
  \frac{\rmd \mathbf{p}}{\rmd t} = -M \cdot \mathbf{p}(t),
\end{equation}
with {\it generator matrix} $M$:
\begin{equation}\label{eq:CRW generator}
  M_{ij} = \begin{cases} 
      -\gamma A_{ij}, & i\neq j, \\
      \gamma\, \mathrm{outDeg}(j), & i = j.
    \end{cases}
\end{equation}
The parameter $\gamma>0$ determines the CRW transition rate between vertices.
The off-diagonal elements of $M_{ij}$ represent the individual probability flows
$j\rightarrow i$ along each edge from vertex $j$, while the diagonal elements
$M_{jj}$ account for the total outflow from vertex $j$ per unit time.

The solution of Eq.~(\ref{eq:CRW master}) can be expressed as a matrix
exponential:
\begin{equation}\label{eq:CRW matrix exp}
  \mathbf{p}(t) = \rme^{-M t} \cdot \mathbf{p}(0).
\end{equation}

\subsection{Quantum walks}
\label{subsection:QW}

A QW may be defined in a natural way by analogy to the CRW. The graph structure
of $G$ is mapped onto a quantum mechanical Hilbert space in which the set of
vertices forms an orthonormal basis $\{\ket{1},\ldots,\ket{N}\}$. The
probability vector $\mathbf{p}(t)$ from the CRW is replaced by a quantum state
vector of probability amplitudes:
\begin{equation}\label{eq:psi definition}
  \ket{\psi(t)} = \sum_{i=1}^{N}\ket{i}\braket{i}{\psi(t)}.
\end{equation}
The amplitudes $\braket{i}{\psi(t)}$ represent a coherent superposition over all
vertices, such that the probability associated with vertex $i$ at time $t$ is
$|\braket{i}{\psi(t)}|^2$. The CRW master equation (Eq.~\ref{eq:CRW master}) is
replaced by a Schr{\"o}dinger equation:
\begin{equation}\label{eq:schrodinger}
  \frac{\rmd \ket{\psi(t)}}{\rmd t} = -\rmi \hat{H} \ket{\psi(t)},
\end{equation}
where $\hat{H}$ is the Hamiltonian, whose matrix elements come from the CRW
generator matrix:
\begin{equation}\label{eq:QW hamiltonian}
  \bra{i}\hat{H}\ket{j} = M_{ij}.
\end{equation}
In the last two equations we have introduced the $\hat{}$ notation to denote
that we regard $\hat{H}$ as an {\it operator}, independent of any
basis. However, in the remainder of this paper we will switch freely between the
operator ($\hat{H}$) and matrix ($H$) viewpoints depending on the context; the
distinction is generally not significant.

The imaginary $\rmi$ factor on the right hand side of Eq.~(\ref{eq:schrodinger})
is responsible for the uniquely ``quantum'' properties of the QW. It should be
noted that, in order for the time evolution of the QW to be unitary (and hence
probability-conserving), $\hat{H}$ must be Hermitian. From Eqs.~(\ref{eq:QW
  hamiltonian}) and (\ref{eq:CRW generator}), it follows that $M$ must
be symmetric, and hence the graph $G$ must be undirected for a QW.

The solution to Eq.~(\ref{eq:schrodinger}) can once again be expressed as a
matrix exponential:
\begin{equation}\label{eq:QW matrix exp}
  \ket{\psi(t)} = \rme^{-\rmi \hat{H} t} \ket{\psi(0)},
\end{equation}
with the initial condition $\ket{\psi(0)} = \ket{q}$ for some $q$.

\subsection{Quantum stochastic walks}
\label{subsection:QSW}

The QW just described is restricted to undirected graphs on which the
Hamiltonian operator is Hermitian. It is also, by definition, limited to pure
quantum states undergoing phase-coherent unitary evolution. By contrast, the CRW
represents the opposite limit, in which scattering is non-unitary and evolution
is purely phase-incoherent. The QSW was proposed by Whitfield \etal
\cite{whitfield2010} as a generalization that includes both QWs and CRWs as
limiting special cases, and allows random walks with a combination of coherent
and incoherent dynamics. The incoherent part of the QSW can be used to
incorporate scattering along directed edges and thus allows directed graphs to
be treated.

In place of the probability vector $\mathbf{p}(t)$ of the CRW or wavefunction
$\ket{\psi(t)}$ of the QW, the QSW is framed in terms of the density matrix
$\hat{\rho}(t)$ \cite{nielsen2010, bellac2011, sakurai2013}. In the general
case, this operator describes a statistical ensemble of pure quantum states
(a so-called ``mixed'' state):
\begin{equation}\label{eq:density matrix}
  \hat{\rho}(t) = \sum_{k} p_k \ket{\psi_k(t)}\bra{\psi_k(t)}.
\end{equation}
The weights $p_k\ge 0$ satisfy $\sum_k p_k=1$ and represent the probabilities for
the system to be in each of the quantum states $\ket{\psi_k(t)}$. This probability
reflects a lack of knowledge of the true system state, which is distinct from
the quantum mechanical uncertainty associated with the measurement a single
quantum state that happens to be a superposition of observable basis states. The
special case of a known quantum state is recovered when only one of the $p_k$ is
non-zero, and is known as a ``pure'' state.

We note several relevant properties of $\hat{\rho}(t)$ which follow from
Eq.~(\ref{eq:density matrix}) and are relevant here: $\hat{\rho}(t)$ is
Hermitian; $\mathrm{tr}(\hat{\rho}(t))=1$ and $\mathrm{tr}(\hat{\rho}(t)^2)\le
1$, with equality holding only for pure states; the expectation value of an
operator $\hat{A}$ is given by $\langle\hat{A}\rangle =
\mathrm{tr}(\hat{\rho}(t)\hat{A})$.

For a QSW on graph $G$, the natural representation for $\hat{\rho}(t)$ is a
$N\times N$ matrix in the basis of vertex states $\{\ket{1},\ldots,\ket{N}\}$,
with elements $\rho_{ij}(t)=\bra{i}\hat{\rho}(t)\ket{j}$. As for the QW, the
usual initial condition for a QSW will have the walker start in the state
$\ket{q}$ at $t=0$, so that $\hat{\rho}(0) = \ket{q}\bra{q}$, or equivalently,
$\rho_{ij}(0)=\delta_{iq}\delta_{jq}$. For $t\ge0$, the diagonal elements
$\rho_{ii}(t)$ represent the probability density at vertex $i$ (and are
therefore referred to as ``populations''), while the off-diagonal elements
$\rho_{ij}(t)$ ($i\neq j$) describe the phase coherence between distinct
vertices $i$ and $j$ (and are known as ``coherences'').

Following Ref.~\cite{whitfield2010}, we can now define the QSW by the following
master equation:
\begin{multline} \label{eq:QSW master}
  \frac{\rmd\hat{\rho}}{\rmd t}
  = -(1-\omega) \rmi [\hat{H},\hat{\rho}(t)] + \\
  \omega \sum_{k=1}^K \left(
  \hat{L}_k \hat{\rho}(t) \hat{L}_k^\dagger - \frac{1}{2}\left(
  \hat{L}_k^\dagger\hat{L}_k\hat{\rho}(t) +
  \hat{\rho}(t)\hat{L}_k^\dagger\hat{L}_k
  \right)\right),
\end{multline}
where $\hat{H}$ is the Hamiltonian operator, describing coherent evolution;
$\hat{L}_k$ are the {\it Lindblad operators}, which
describe phase-incoherent scattering (see below); and $0\le\omega\le1$ is a
weighting factor that interpolates between the coherent and incoherent terms.

Eq.~(\ref{eq:QSW master}) is an example of the Kossakowski-Lindblad master
equation \cite{kossakowski1972, lindblad1976}, widely used to model interactions
with an external environment in the study of open quantum systems
\cite{breuer2002, rivas2011}. In its usual context the $1-\omega$ and $\omega$
factors are not present, but it is convenient to include them in the QSW as they
permit an explicit variation between completely phase coherent ($\omega=0$) and
incoherent ($\omega=1$) regimes.

The first term on the right hand side of Eq.~(\ref{eq:QSW master}), representing
the phase coherent component of the evolution,
uses the same Hamiltonian $\hat{H}$ that we introduced in the QW. Indeed,
when $\omega=0$, Eq.~(\ref{eq:QSW master}) reduces to the Liouville-von
Neumann equation, which is the density matrix equivalent of the Schr\"odinger
equation Eq.~(\ref{eq:schrodinger}).

As for the QW, $\hat{H}$ must be Hermitian and hence based on an undirected
graph. Unlike the QW case, however, in a QSW the graph $G$ may be undirected in
general. We therefore define $\hat{H}$ in terms of a symmetrized version of $G$,
as follows:
\begin{equation} \label{eq:QSW hamiltonian}
  H_{ij} = \begin{cases}
  -\gamma\, \mathrm{max}(A_{ij},A_{ji}), & i\neq j, \\
  -\sum_{k\neq j}H_{kj}, & i = j. 
  \end{cases}
\end{equation}

The directedness of $G$ is manifest in the second term of Eq.~(\ref{eq:QSW master}), which represents the incoherent component
of the evolution. It comprises a sum over the Lindblad operators $\hat{L}_k$,
each of which corresponds to a particular scattering channel. The range of the sum, $K$, depends on the particular choice of $\{\hat{L}_k\}$. For a QSW, a natural choice is the set of operators representing scattering between each pair of vertices. Identifying $k$ with the
pair $i,j$ (say, $k \equiv N(j-1)+i$) we define
\begin{equation} \label{eq:Lk definition}
  \hat{L}_k = \sqrt{|M_{ij}|}\ket{i}\bra{j},
\end{equation}
representing incoherent scattering from vertex $j$ to $i$. The sum extends over
all $i,j$ so that $K = N^2$, but only pairs with non-zero $M_{ij}$ contribute.
In this way, the scattering represented by the $\hat{L}_k$ operators
incorporates the full directed structure of $G$; this is what is meant by saying
that a QSW can be applied to a directed graph. It is straightforward to show
that with this set of $\hat{L}_k$, the QSW reduces to the CRW
when $\omega=1$, if we make the obvious identification $\rho_{ii}(t)\equiv
p_i(t)$.

Other choices for $\hat{L}_k$ are possible and may lead to QSWs with different
physical interpretations. However, the definition in Eq.~(\ref{eq:Lk
  definition}) is of particular interest, since it allows us to recover the
well-known and widely studied QW ($\omega=0$) and CRW ($\omega=1$) as special
cases. By varying $\omega$ between 0 and 1 one can then explore intermediate
regimes with both quantum and classical behaviour.

\section{The QSWalk package}
\label{section:package}

The {\tt QSWalk} package is written in the {\it Mathematica} system (also known
since 2013 as the Wolfram Language) \cite{wolfram2016}. One of the key strengths
of {\it Mathematica} is its vast number of built-in functions (approximately
5000 as of version 10.3), providing efficient and sophisticated numerical and symbolic algorithms.

Of particular value to our application is {\it Mathematica}'s built-in graph
theory functionality: graphs are ``first-class citizens'' \cite{wolframURL} and
many standard graph types are available as built-in functions. This makes it
easy to apply our package to immediately study a wide range of known graph
types, in addition to readily defining new ones.

{\it Mathematica} also offers powerful linear algebra routines, which are
essential for the efficient evaluation of QSWs. In particular, we make use of
the built-in sparse array methods for the matrix exponential calculation. The
use of these methods is seamless and mostly transparent to the user, which
greatly streamlines the development of efficient code.

\subsection{Package overview}

The main purpose of the {\tt QSWalk} package is the implementation of QSWs on
graphs. In this section we illustrate this generic use case with a simple
example that nevertheless incorporates all the necessary steps. For ease of
presentation the following input and output is taken from a command-line
terminal session (although we recommend working within the {\it Mathematica} notebook interface in
general).

\begin{verbatim}
Mathematica 10.3.0 for Mac OS X x86 (64-bit)
Copyright 1988-2015 Wolfram Research, Inc.
\end{verbatim}
We begin by loading the package. For this to work, the {\tt QSWalk.m} file must be in
the current path (as specified by the {\tt \$Path} variable). Alternatively, the
full path to the file can be specified.
\begin{verbatim}
In[1]:= <<QSWalk.m
\end{verbatim}
Our calculation begins by defining a graph object; here we use a built-in
function to create an Erd{\"o}s-Renyi random graph with 10 vertices and 30
(directed) edges:
\begin{verbatim}
In[2]:= G = RandomGraph[{10,30}, DirectedEdges->True];
\end{verbatim}
Assuming a transition rate $\gamma=1$, we define the Hamiltonian as the
generator matrix of the symmetrized graph (Eq.~\ref{eq:QSW hamiltonian}):
\begin{verbatim}
In[3]:= gamma = 1.;

In[4]:= H = GeneratorMatrix[UndirectedGraph[G], gamma];
\end{verbatim}
We then define the Lindblad operators using the generator of the original
directed graph (Eq.~\ref{eq:Lk definition}):
\begin{verbatim}
In[5]:= LkSet = LindbladSet[GeneratorMatrix[G,gamma]];
\end{verbatim}
Next, we define the remaining inputs. The initial density
matrix is $\hat{\rho}(0) = \ket{1}\bra{1}$, which is conveniently implemented in
matrix form using the {\tt SparseArray} function:
\begin{verbatim}
In[6]:= omega = 0.5; t = 10.;

In[7]:= rho0 = SparseArray[{{1,1}->1.}, {n,n}];
\end{verbatim}
The final step is to carry out the QSW calculation:
\begin{verbatim}
In[8]:= rho1 = QuantumStochasticWalk[H,LkSet,omega,rho0,t];
\end{verbatim}
The output is a density matrix; to get the final vertex probabilities we take
the diagonal elements:
\begin{verbatim}
In[8]:= p = Diagonal[rho1] // Chop

Out[8]= {0.311295, 0.0266081, 0.0938145, 0.0797414, 0.0884782,

>  0.0687126, 0.0747984, 0.121446, 0.0900074, 0.0450989}
\end{verbatim}
The above eight short lines of code can, with only slight modifications, be used
to calculate QSWs on arbitrary graphs.

It is often of interest to evaluate a walk at a sequence of times, with the
result of each time step forming the input to the next step. In this case,
it is advantageous to avoid re-generating the master equation matrix for each
time step. The following approach shows how this can be done in a
straightforward way.

First, define a time increment {\tt dt}. We then call {\tt
  QuantumStochasticWalk} as before, but with a symbolic (\ie undefined) initial
state {\tt rho}. The result is a partially evaluated expression in which the
master equation has been generated, but the matrix exponential has remained
unevaluated because its second argument is symbolic. We save this as a function
{\tt qsw}, which can subsequently be called with specific numerical values for {\tt rho}:
\begin{verbatim}
In[12]:= dt = 1.;

In[13]:= qsw[rho_]=QuantumStochasticWalk[H,LkSet,omega,rho,dt]                                                                

Out[13]= Matricize[MatrixExp[SparseArray[<1170>, {100, 100}],
>  Vectorize[rho]]]
\end{verbatim}
We use the {\tt NestList} function to call {\tt qsw} recursively, starting
with {\tt rho = rho0} to build up a list of density matrices for each time:
\begin{verbatim}
In[14]:= pt = Diagonal /@ NestList[qsw,rho0,Round[t/dt]];    
\end{verbatim}
As a check, we can verify that the final element in this list agrees with our
previous calculation:
\begin{verbatim}
In[20]:= Chop[Last[pt] - p]                                                                                                         

Out[20]= {0, 0, 0, 0, 0, 0, 0, 0, 0, 0}
\end{verbatim}
The {\tt Chop} function is used here to zero out the small differences arising
from the limitations of machine precision arithmetic.

\subsection{Functions of the QSWalk package}

In this section we provide a brief summary of the key functions in the {\tt
  QSWalk} package.

\subsubsection*{Propagator functions}

\noindent The function {\tt QuantumStochasticWalk} implements QSWs on graphs and
is the main function in the {\tt QSWalk} package. We also provide functions to
implement QWs and CRWs, which are much faster to evaluate (as they work with
vectors of length $N$ rather than $N^2$) and provide a useful comparison.

\begin{itemize}
\item {\tt QuantumStochasticWalk[H,LkSet,omega,rho0,t]}: computes a QSW with
  Hamiltonian matrix {\tt H}, Lindblad operators {\tt LkSet}, and weighting
  factor {\tt omega}; starting from density matrix {\tt rho0} at time 0 and
  returning the density matrix at time {\tt t}. {\tt H} should be an $N\times N$
  Hermitian matrix, and {\tt LkSet} should be a list of $N\times N$ matrices;
  {\tt rho0} should be an $N\times N$ Hermitian matrix with a trace of unity.

\item {\tt QuantumWalk[H,psi0,t]}: computes a QW with Hamiltonian matrix {\tt
  H}, starting from state vector {\tt psi0} at time 0 and returning the state
  vector at time {\tt t}. {\tt H} should be an $N\times N$ Hermitian matrix;
  {\tt psi0} should be a length-$N$ list of complex-valued probability
  amplitudes with squared magnitudes summing to unity.

\item {\tt ClassicalRandomWalk[M,p0,t]}: computes a CRW with generator matrix
  {\tt M}, starting from probability vector {\tt p0} at time 0 and returning the
  probability vector at time {\tt t}.  {\tt M} should be a real-valued $N\times
  N$ matrix; {\tt p0} should be a length-N list of non-negative probabilities
  summing to unity.
\end{itemize}

\subsubsection*{Constructing Hamiltonian and Lindblad operators}

\noindent These functions are used to construct the inputs to the
functions from the previous section (\ie {\tt H}, {\tt M}, and {\tt LkSet}).

\begin{itemize}
\item {\tt GeneratorMatrix[G,gamma]}: returns the generator matrix $M$ for the
  graph {\tt G}, with transition rate {\tt gamma}.

\item {\tt LindbladSet[mat]}: returns of set of Lindblad matrices corresponding
  to each non-zero element of {\tt mat}.

\item {\tt GoogleMatrix[G,alpha]}: returns the ``Google matrix'' with damping
  factor alpha for the graph {\tt G}.
\end{itemize}

\subsubsection*{Vectorization functions}

\noindent The following two functions convert between an $N\times N$ matrix and
its $N\times1$ vectorized form (see Section \ref{section:vectorization}).

\begin{itemize}
\item {\tt Vectorize[mat]}: returns the vectorization of matrix {\tt mat}.

\item {\tt Matricize[vec]}: returns the matrix formed by applying the inverse
  action of the {\tt Vectorize} function on {\tt vec}. The length of {\tt vec}
  must be an integer squared.
\end{itemize}

\subsubsection*{Graph functions}

\noindent Here we provide definitions for several graph types that have been
used in studies on quantum walks, but are not already available in {\it
  Mathematica}.

\begin{itemize}
\item {\tt CayleyTree[d,n]}: returns a graph representing an $n$-th generation
  Cayley tree of order $d$ \cite{ostilli2012}. QWs have been applied to such
  graphs in the study of quantum search algorithms \cite{agliari2010,
    berry2010}.

\item {\tt GluedBinaryTree[n]}: returns a graph comprising two complete,
  $(n+1)$-level binary trees glued together in order along their leaf
  nodes. This graph type was used by Childs \etal as an example for which QWs
  provide an exponential speedup over a CRW \cite{childs2002}.

\item {\tt RandomGluedBinaryTree[n]} returns a graph comprising two complete,
  $(n+1)$-level binary trees glued together in random order along their leaf
  nodes. These were important in demonstrating the existence of graphs for which
  no classical algorithm could compete with a quantum walk \cite{childs2003}.
\end{itemize}

\subsection{Vectorizing the master equation}
\label{section:vectorization}

We now describe our approach for solving the QSW master equation
Eq.~(\ref{eq:QSW master}) in the {\tt QSWalk} package. For clarity, in this
section we adopt the matrix viewpoint and drop the $\hat{}$ notation throughout.

The master equation is a superoperator acting on the density matrix
$\rho(t)$. In this form, a solution to Eq.~(\ref{eq:QSW master}) using the
matrix exponential, by analogy to Eqs.~(\ref{eq:CRW matrix exp}, \ref{eq:QW
  matrix exp}), cannot be found.  We can, however, recast $\rho(t)$ as the
$N^2$-element column vector $\tilde{\rho}(t)$, formed by concatenating its
columns in order:
\begin{equation}\label{eq:vectorized rho}
  \tilde{\rho}(t) =
  (\rho_{11}(t),\ldots,\rho_{N1}(t),\rho_{12}(t),\ldots,\rho_{N2}(t),
  \ldots,\rho_{1N}(t),\ldots,\rho_{NN}(t))^T.
\end{equation}
This operation is known in linear algebra as {\it vectorization}
\cite{banerjee2014}. In vectorized form, Eq.~(\ref{eq:QSW master}) reduces to a
linear operator (albeit in an expanded $N^2\times N^2$ space), whose solution
{\it can} be expressed as a matrix exponential.

\begin{figure}\label{fig:arrays}
  \centering
  \includegraphics[width=0.4\textwidth]{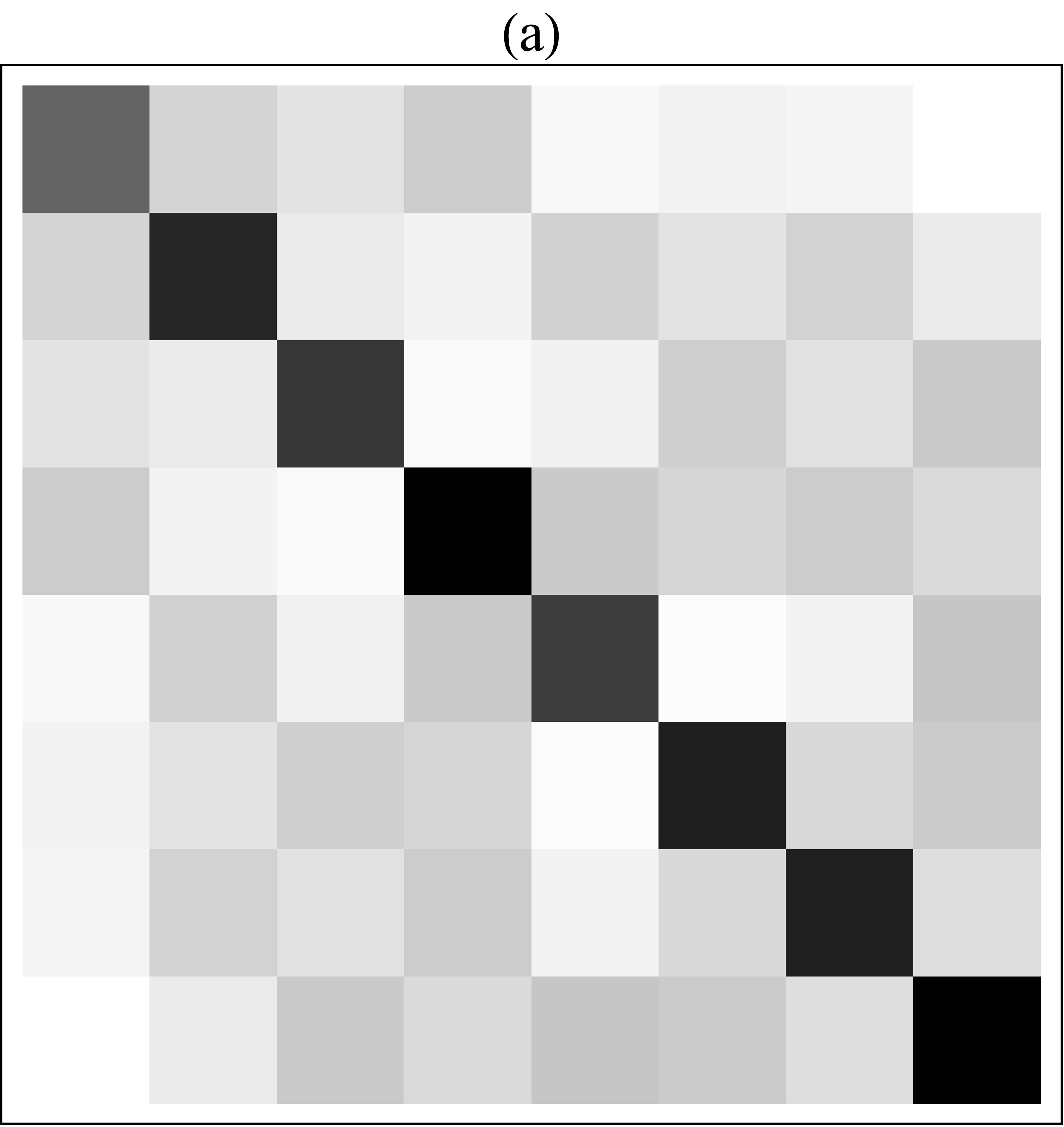}
  \hspace{0.1\textwidth}
  \includegraphics[width=0.4\textwidth]{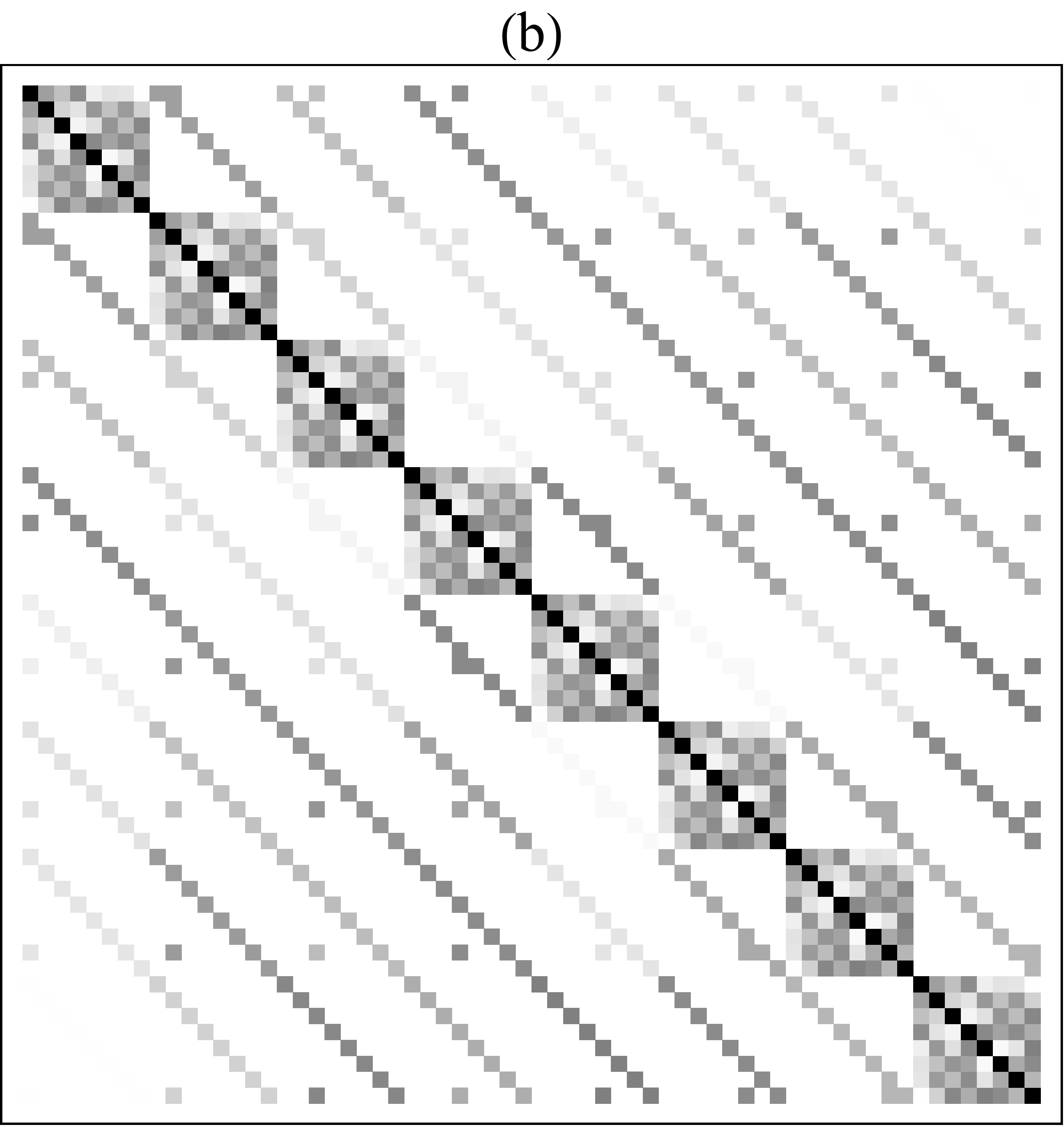}
  \caption{(a) Hamiltonian matrix of a complete undirected graph with $N=8$ and
    random edge weights between 0 and 1; (b) the corresponding sparse matrix
    $\mathcal{L}$ (Eq.~\ref{eq:QSW master L}). The plots are shaded with
    darkness proportional to the magnitude of matrix entries, so that white
    areas correspond to zero entries.}
\end{figure}

To vectorize Eq.~(\ref{eq:QSW master}) we make use of a standard identity
(\cite{banerjee2014}, Theorem 14.14) for arbitrary $N\times N$ matrices $X,Y,Z$:
\begin{equation}\label{eq:vectorization identity}
  \mathrm{vec}(X\cdot Y \cdot Z) = (Z^T \otimes X) \cdot \mathrm{vec}(Y), 
\end{equation}
where $\otimes$ denotes the Kronecker (direct) product, ${}^T$ is the matrix
transpose, and $\cdot$ is standard matrix multiplication. The resulting
vectorized QSW master equation is:
\begin{equation}\label{eq:QSW master vectorized}
  \frac{\rmd \tilde{\rho}}{\rmd t} = \mathcal{L}\cdot\tilde{\rho}(t)
\end{equation}
where
\begin{multline} \label{eq:QSW master L}
  \mathcal{L} =
  -(1-\omega)\rmi \left(I_N\otimes H - H^T\otimes I_N\right) + \\
  \omega \sum_{k=1}^{K}\left(
  L_k^{\ast}\otimes L_k - \frac{1}{2}\left(
  I_N\otimes L_k^\dagger L_k + L_k^TL_k^{\ast}\otimes I_N
  \right)\right).
\end{multline}
The value of this representation is that each term is a direct product involving
one or more sparse matrices ($I_N$ and $L_k$), and therefore $\mathcal{L}$ is a
sparse matrix. Fig.~\ref{fig:arrays} illustrates this point for a complete
graph, for which $H$ is dense. 

The matrix exponential solution of Eq.~(\ref{eq:QSW master vectorized}) is then
\begin{equation}\label{eq:QSW matrix exp}
  \tilde{\rho}(t) = \rme^{\mathcal{L} t}\cdot\tilde{\rho}(0). 
\end{equation}
The density matrix $\rho(t)$ is trivially recovered from the vector
$\tilde{\rho}(t)$.

Eq.~(\ref{eq:QSW master vectorized}) can be implemented using {\it
  Mathematica}'s built-in functions; in particular, {\tt KroneckerProduct}, {\tt
  IdentityMatrix}, {\tt Conjugate}, {\tt Transpose}, {\tt ConjugateTranspose},
as well as matrix multiplication (the ``{\tt .}'' operator).

The vectorization approach outlined above leads us to work with matrices of size
$N^2\times N^2$, which quickly become rather large for even moderately-sized
graphs. It therefore becomes essential to exploit sparse matrix
representations. In {\it Mathematica}, sparse arrays are implemented via the
{\tt SparseArray} function \cite{wolframURLa}, and all of the built-in matrix
functionality includes handling to work with {\tt SparseArray} objects
efficiently. Therefore, provided we correctly initialize all the relevant
matrices to be in {\tt SparseArray} form, {\it Mathematica} will automatically
carry out all matrix operations using sparse matrix methods if possible.

\begin{figure}\label{fig:scaling}
\centering
\includegraphics[width=0.45\textwidth]{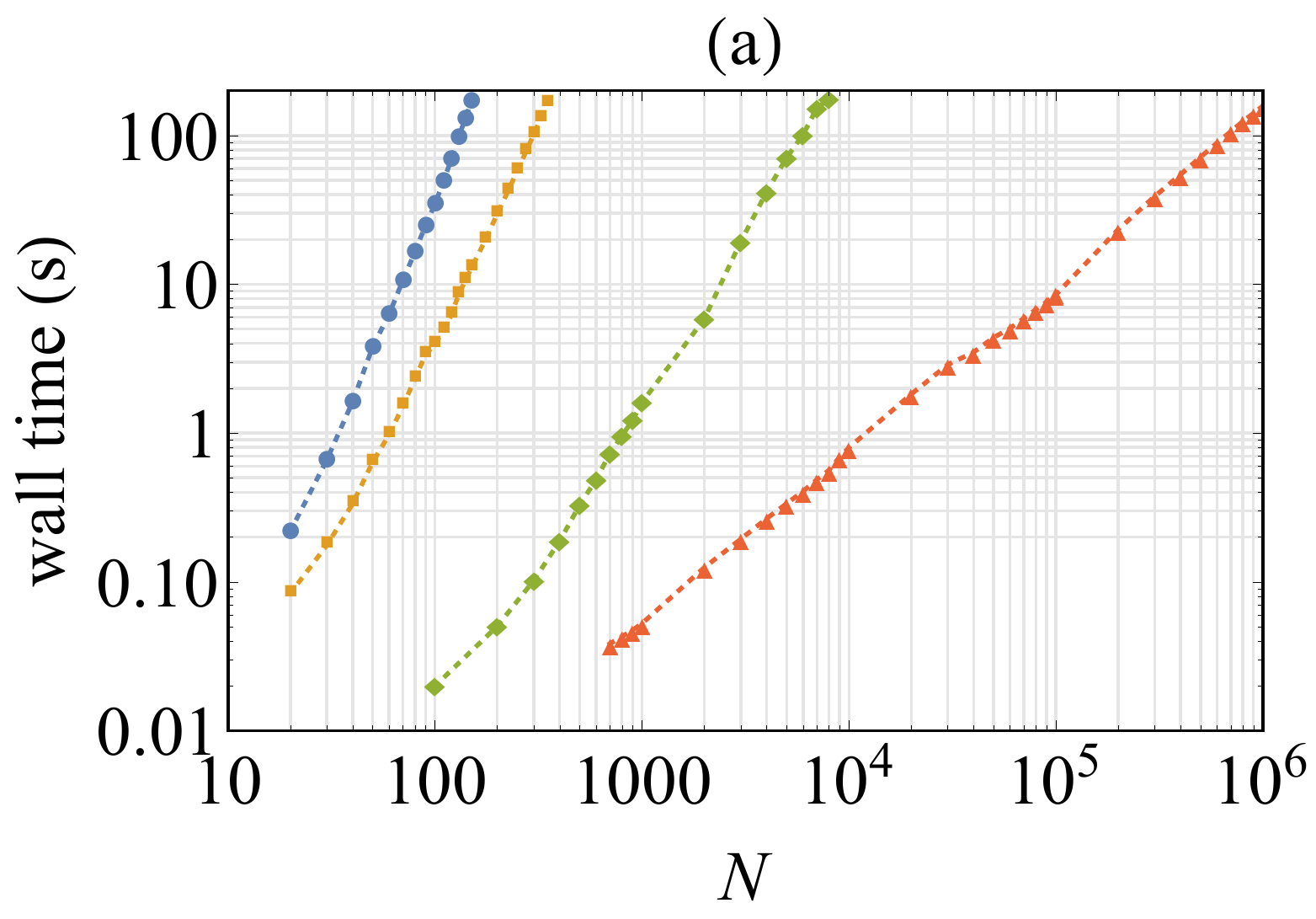}
\hspace{0.05\textwidth}
\includegraphics[width=0.45\textwidth]{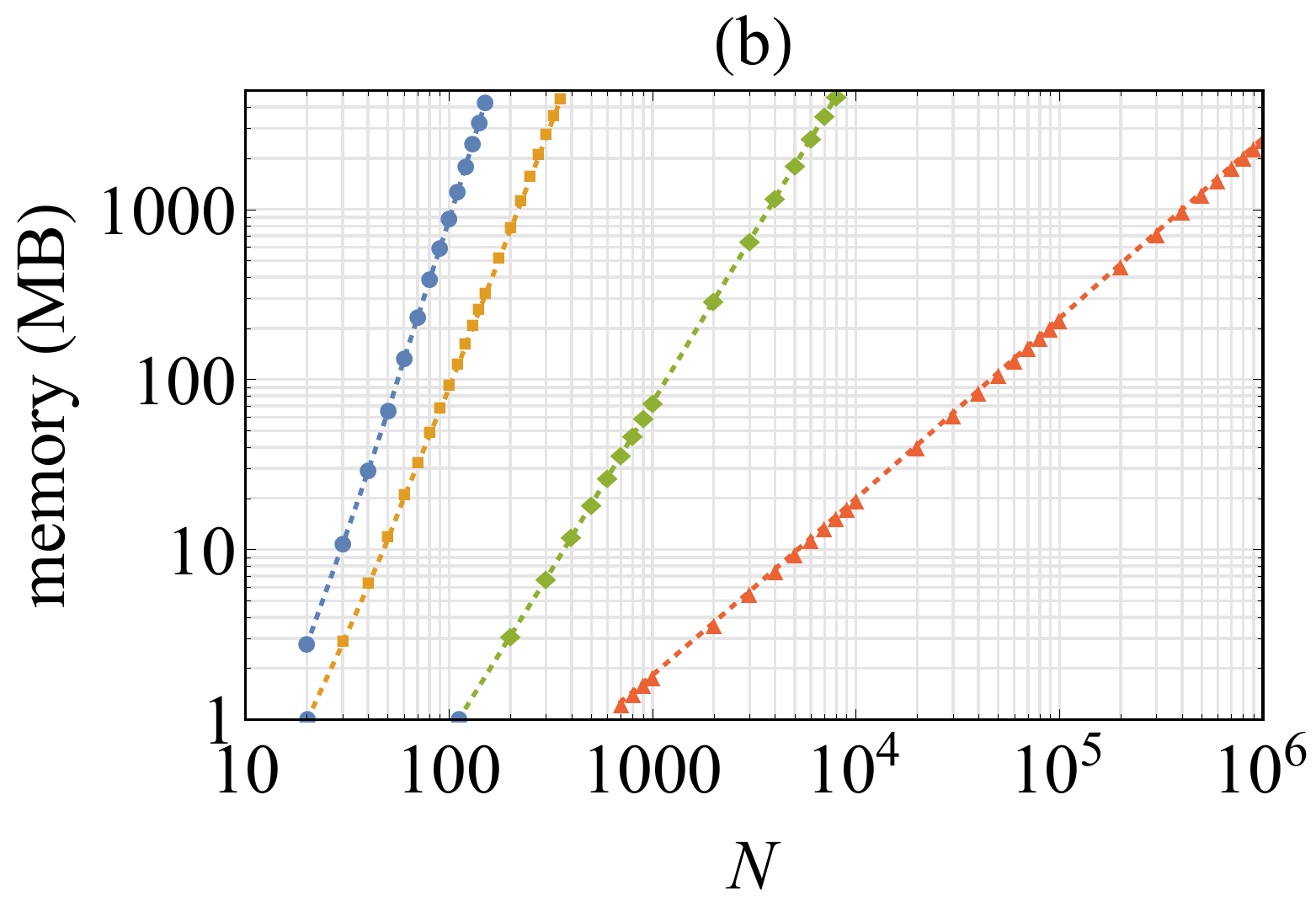}
\includegraphics[width=0.45\textwidth]{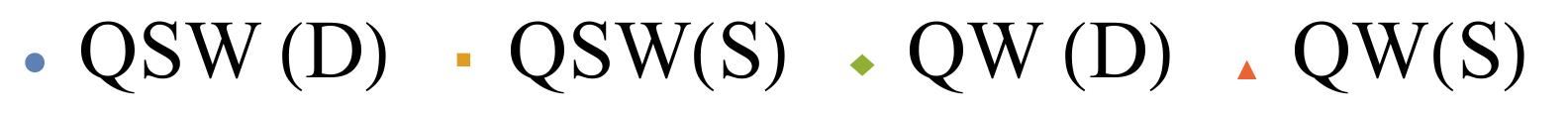}
\caption{Scaling behaviour for QSWs and QWs as a function of graph size $N$, for
  representative examples of dense (D) and sparse (S) undirected graphs. The
  dense graphs have $N(N-1)/2$ edges (defined with {\tt CompleteGraph}), while
  the sparse graphs are of Erd\"os-Renyi type with $\sim N \mathrm{log}N$ edges
  (defined with {\tt RandomGraph}); in both cases the edges are assigned random
  weights uniformly between 0 and 1. In these figures, the slope of the lines
  gives the scaling exponent with respect to $N$. In (a), these are
  approximately (in the same order as the legend): 3.3, 2.7, 2.2, and 1.1; in
  (b), they are 3.7, 3.0, 2.0, and 1.1. The parameters used are $\gamma=1$,
  $\omega=0.5$, $t=10$, with initial conditions $\rho_{ij}(0)=\delta_{ij}/n$
  (QSW) and $\braket{i}{\psi(0)}=1/\sqrt{n}$ (QW). Calculations were performed
  on a Late-2013 iMac with 8GB RAM, running Mathematica 10.3 on Mac OS 10.11.}
\end{figure}

The performance of our approach depends on both the size (number of vertices)
and density (number of edges) of the graph being studied. Fig.~\ref{fig:scaling}
gives some indicative estimates of running time and memory requirements for
representative dense and sparse graphs. It can be seen that the primary
bottleneck on a typical machine is memory: typically, memory requirements for a
QSW can be as high as 4GB with run times of only several minutes. This
corresponds to $N\approx150$ for dense graphs, and $N\approx350$ for sparse
graphs. For QWs, unsurprisingly, the scaling is much better: for comparable run
time and memory we can calculate QWs on graphs with $N\approx 10^4$ (dense) and
$N\approx 10^6$ (sparse).

\section{Applications}
\label{section:applications}

In this section we describe several applications of the {\tt QSWalk} package to
systems similar to those studied recently in the literature. Our emphasis
is on demonstrating the functionality of the package rather than describing any
new physics.

\subsection{Quantum-classical transition on a line}
\label{subsection:transition}

One of the key properties of the QSW is that it permits an interpolation between
coherent (QW) and incoherent (CRW) dynamics, via the parameter $\omega$. To
illustrate this transition, we consider a QSW on one of the simplest graphs
possible: an undirected $N\times 1$ line graph (which is equivalent to the
one-dimensional tight-binding model studied in electron transport theory
\cite{datta1997}). The adjacency matrix for such a graph has entries
$A_{ij}=\delta_{i,j+1}+\delta_{i,j-1}$.  The matrix elements of the Hamiltonian
and Lindblad operators can be found using Eqs.~(\ref{eq:QSW hamiltonian}) and
(\ref{eq:Lk definition}):
\begin{equation}
  \bra{i}\hat{H}\ket{j} = \gamma\left(\delta_{ij}
  \left(2-\delta_{i,1}-\delta_{iN}\right)
  - \delta_{i,j+1} - \delta_{i,j-1}\right),
\end{equation}
\begin{equation}
  \bra{i}\hat{L}_{k'}\ket{j} = \delta_{i'i}\delta_{j'j}\sqrt{\gamma}
  \left(\delta_{i'j'} \left(2-\delta_{i',1}-\delta_{i'N}\right)
  + \delta_{i',j'+1} + \delta_{i',j'-1}\right),
\end{equation}
where $k'\equiv(i',j')$ and $\hat{L}_{k'}$ is the Lindblad operator
corresponding to the scattering process $\ket{i'}\bra{j'}$. As initial condition
we assume the state is localized in the middle of the line at vertex $q=(N+1)/2$
(assuming odd $N$), so that $\hat{\rho}(0) = \ket{q}\bra{q}$.

\begin{figure}\label{fig:qw crw transition}
\includegraphics[width=0.45\textwidth]{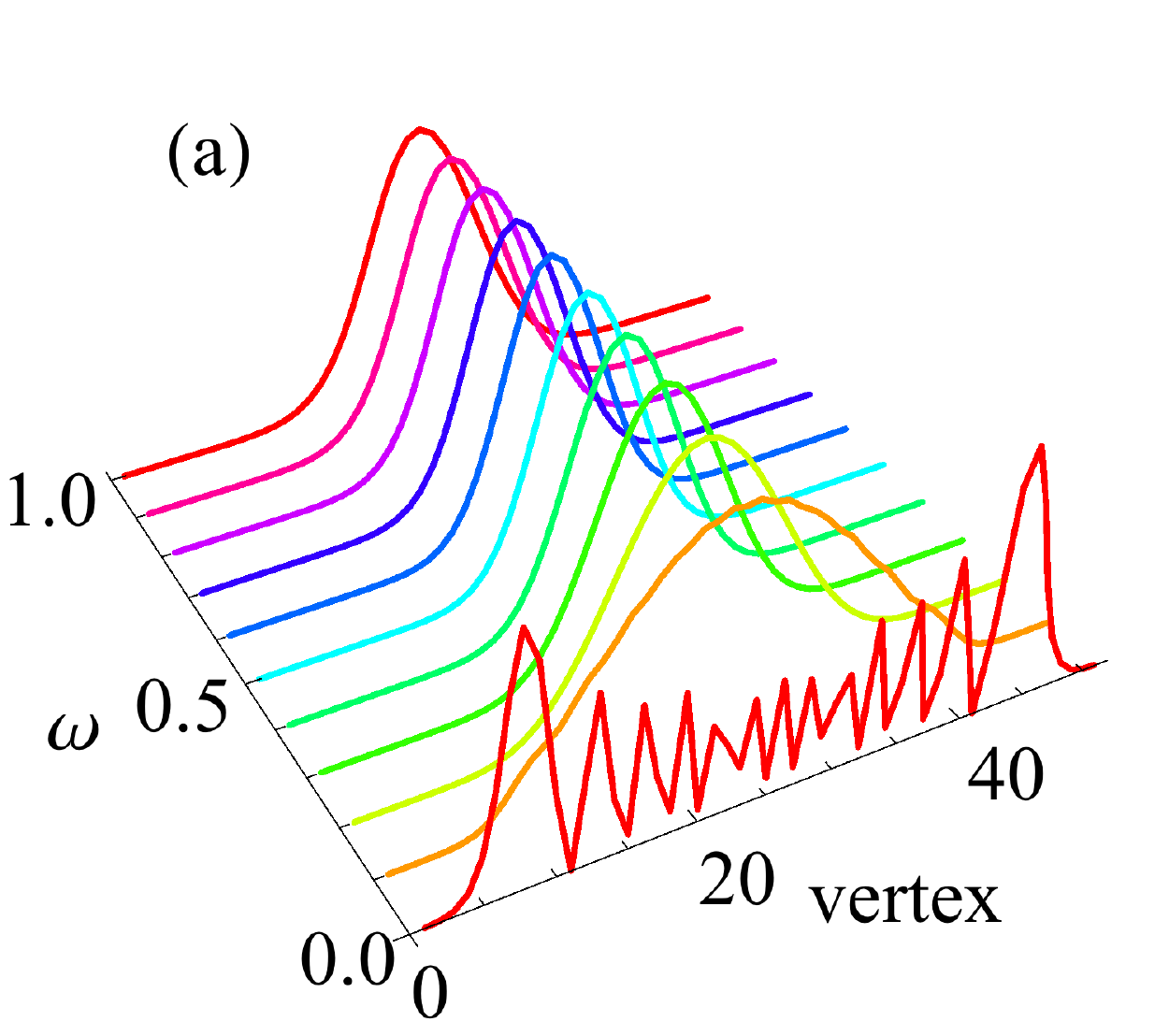}
\hspace{0.05\textwidth}
\includegraphics[width=0.5\textwidth]{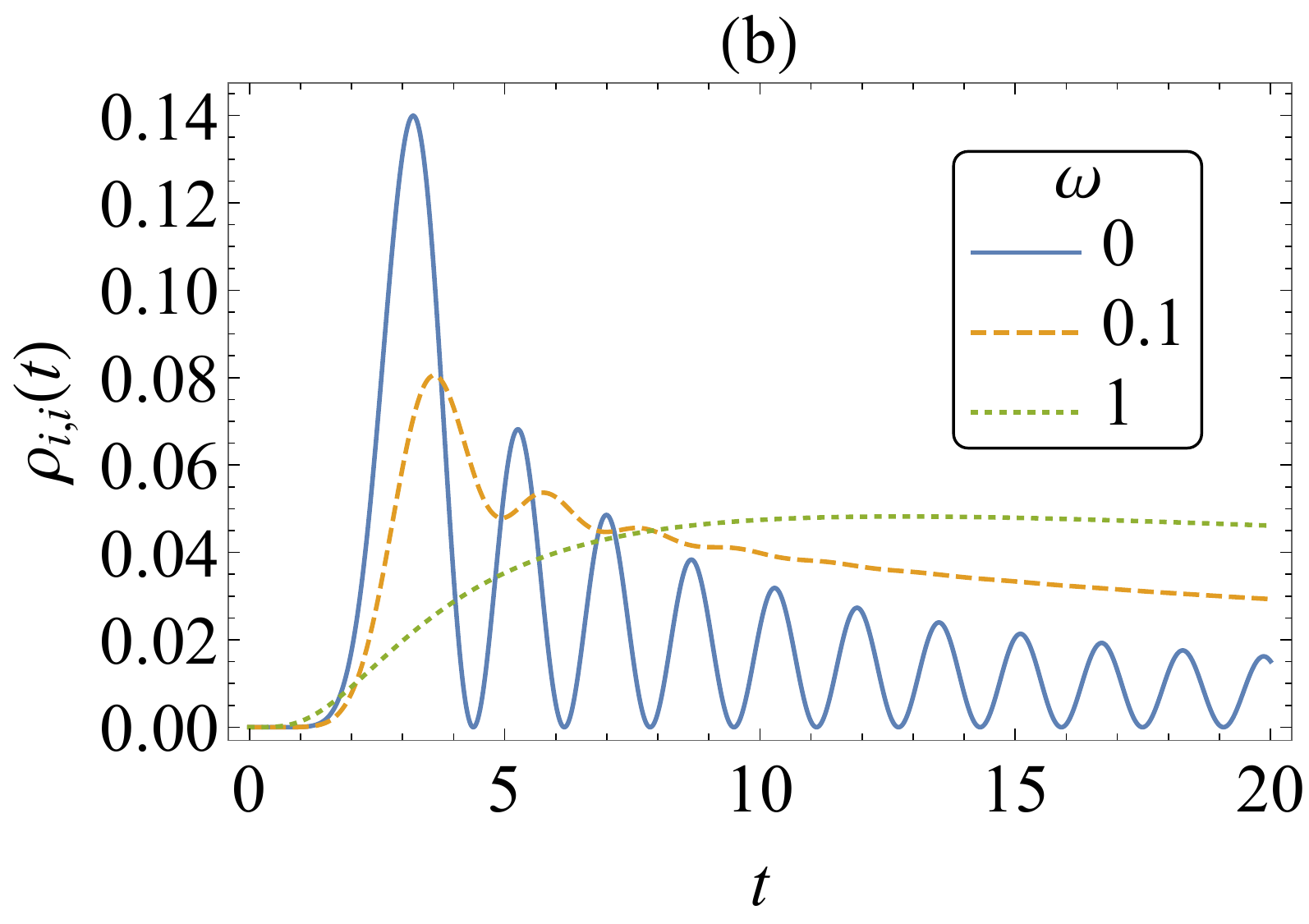}
\caption{(a) A sequence of QSWs showing the transition from pure QW ($\omega=0$)
  to pure CRW ($\omega=1$), on a line graph with $N=51$ vertices and propagation
  time $t=5$ (in atomic units). (b) Time evolution of QSWs on the same graph,
  for a fixed vertex $i=31$ ($\approx0.6N$) for several values of $\omega$.}
\end{figure}

The code to implement this QSW in {\tt QSWalk} is straightforward, and uses the
built-in {\tt GridGraph} function for the line graph:
\begin{verbatim}
n = 51; gamma = 1.; t = 10.; omega = 0.5;
G = GridGraph[{n}];
H = GeneratorMatrix[G, gamma];
LkSet = LindbladSet[H];
rho0 = SparseArray[{{(n+1)/2,(n+1)/2}->1.},{n,n}];
rho1 = QuantumStochasticWalk[H,LkSet,omega,rho0,t];
\end{verbatim}

Fig.~\ref{fig:qw crw transition} shows the dynamics of the resulting QSW. In
Fig.~\ref{fig:qw crw transition}a, the transition from an oscillatory
distribution (characteristic of coherent dynamics) at $\omega=0$, to a purely
diffusive distribution at $\omega=1$, is clearly evident. Fig.~\ref{fig:qw crw transition}b shows the time evolution of a particular vertex population
($i\approx0.6N$) for several values of $\omega$, illustrating the damping which
occurs when $\omega>0$. This is actually a useful property, as it causes the QSW
populations to converge to a stationary state rather than oscillate
indefinitely, while still retaining some (transitory) coherence.

\subsection{Pure dephasing scattering}

As noted in Ref.~\cite{whitfield2010}, an alternative approach to incoherent
scattering is via pure dephasing scattering processes \cite{kendon2007}. In this
model, the Lindblad operators are defined by
\begin{equation} \label{eq:Lk dephasing}
  \hat{L}_k = \ket{k}\bra{k},
\end{equation}
for $k = 1,\ldots,N$.

\begin{figure}\label{fig:dephasing}
\includegraphics[width=0.45\textwidth]{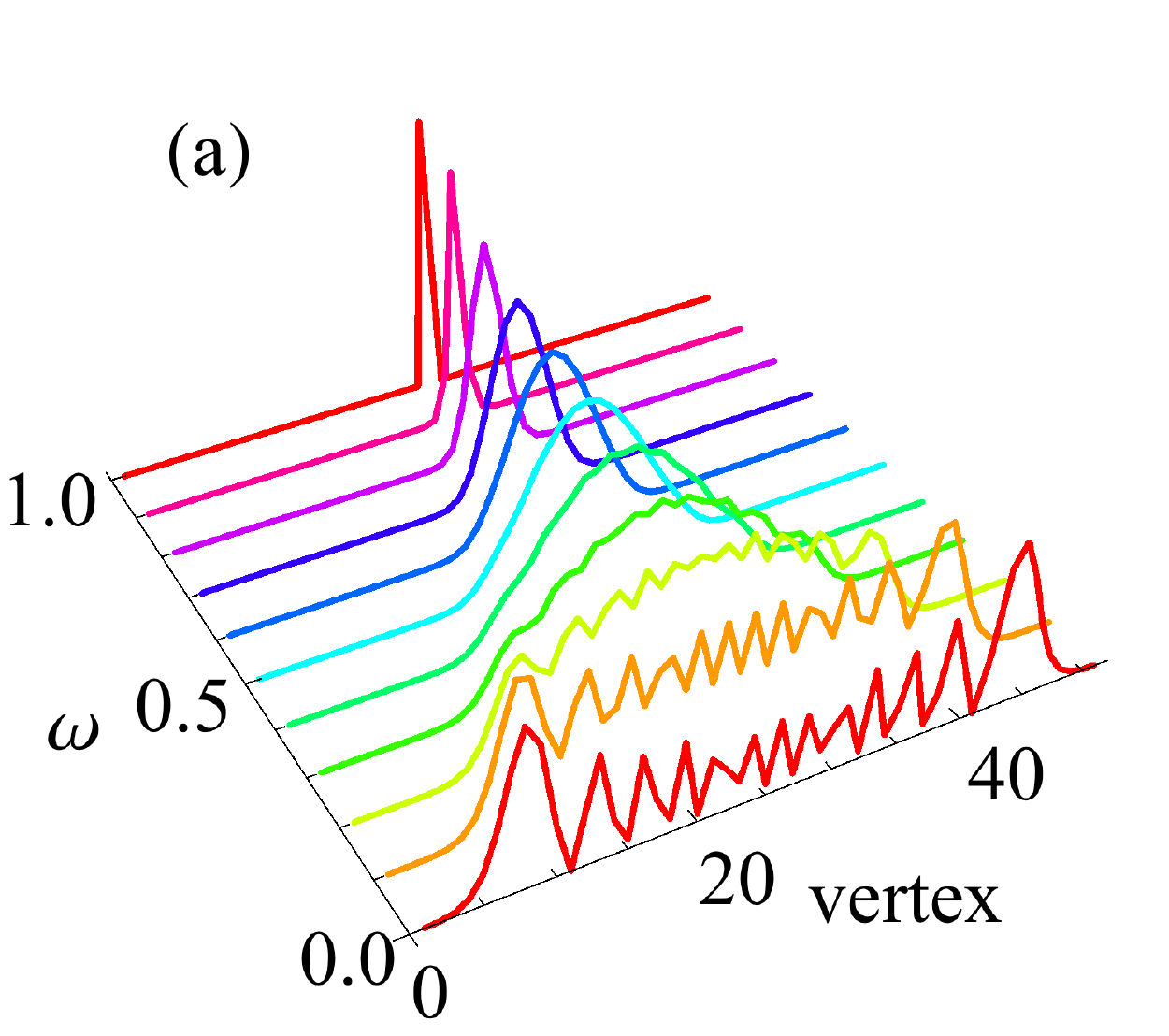}
\hspace{0.05\textwidth}
\includegraphics[width=0.5\textwidth]{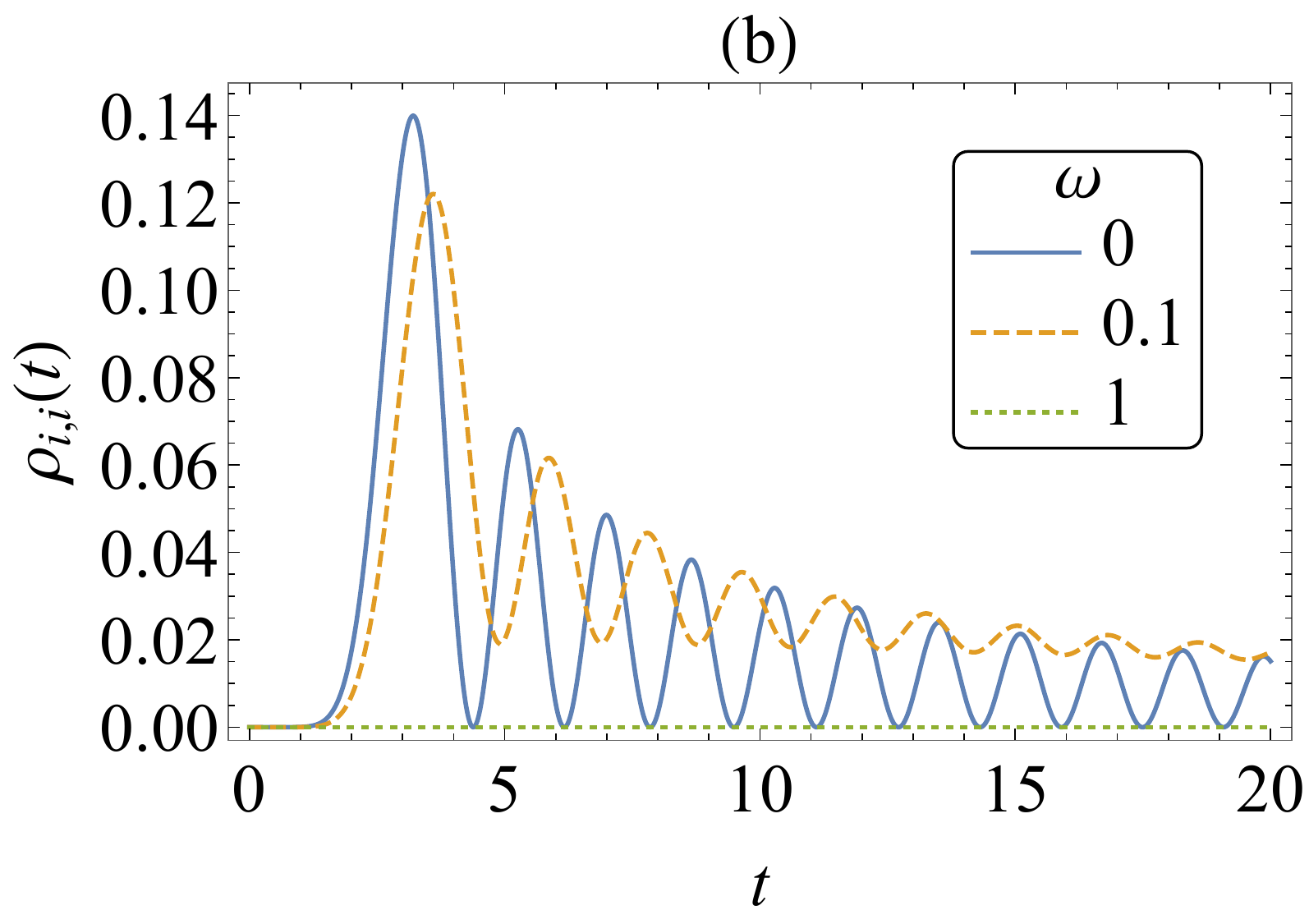}
\caption{The equivalent of the plots from Fig.~\ref{fig:qw crw transition} but
  for a QSW with pure dephasing scattering (Eq.~\ref{eq:Lk dephasing}).}
\end{figure}

The implementation is the same as the previous section, with the exception of
the definition of $\hat{L}_k$, which now reads:
\begin{verbatim}
LkSet = Table[SparseArray[{{i,i}->1}, {n,n}], {i,n}];
\end{verbatim}

In Fig.~\ref{fig:dephasing} we consider a QSW using the pure dephasing model of
Eq.~(\ref{eq:Lk dephasing}) on the line graph from the previous section. There
are several notable differences: firstly, from Fig.~\ref{fig:dephasing}a it is
clear that $\omega\rightarrow1$ no longer produces the CRW. In fact, in this
limit there are no dynamics at all since the $\hat{L}_k$ operators do not
scatter between vertices. In this case, for $\omega=1$ we have
$\hat{\rho}(t) = \hat{\rho}(0)$ for all $t$.

Secondly, comparing Fig.~\ref{fig:dephasing} to Fig.~\ref{fig:qw crw transition}
we see that pure dephasing scattering causes {\it less} dephasing than the model
of Eq.~(\ref{eq:Lk definition}), in the sense that the wave-like oscillations
persist for larger $\omega$. This is actually not surprising since for the line
graph the model of Eq.~(\ref{eq:Lk definition}) has dephasing terms with
weighting $\sqrt{2}$ (in contrast to the weighting of $1$ in Eq.~\ref{eq:Lk
  dephasing}), in addition to the off-diagonal scattering elements which also
contribute to the dephasing.

\subsection{Photosynthetic light-harvesting in the FMO complex}
\label{subsection:fmo}

QSWs have found application to an important problem in biophysics, concerning
the capture of energy (in the form of photons from sunlight) by photosynthetic
protein complexes \cite{blankenship2014}. A prototypical example which has been
extensively studied is the so-called Fenna-Matthews-Olson (FMO) complex from
green sulphur bacteria. This molecule is comprised of seven regions called
chromophores, through which electronic excitations (excitons) are propagated by
a hopping process in which phase coherence plays an important role
\cite{engel2007}. An excitation generated by photon absorption on a particular
chromophore is transmitted through the complex to a reaction centre where it
is converted to stored chemical energy.

Theoretical studies aimed at understanding the phase coherence properties in the
FMO have modeled the chromophores as vertices in an undirected graph on which
excitons follow a quantum walk-like process \cite{mohseni2008, hoyer2010,
  plenio2008} which, like the QSW, combines coherent and incoherent evolution
through a master equation. These models are not strictly equivalent to QSWs as
they incorporate a loss term, representing absorption of the exciton into the
reaction centre, and therefore are not probability-conserving. Nevertheless, it
is straightforward to develop a QSW model which describes the same essential
physics, as we now show.

\begin{figure} \label{fig:fmo}
  \includegraphics[width=0.4\textwidth,valign=c]{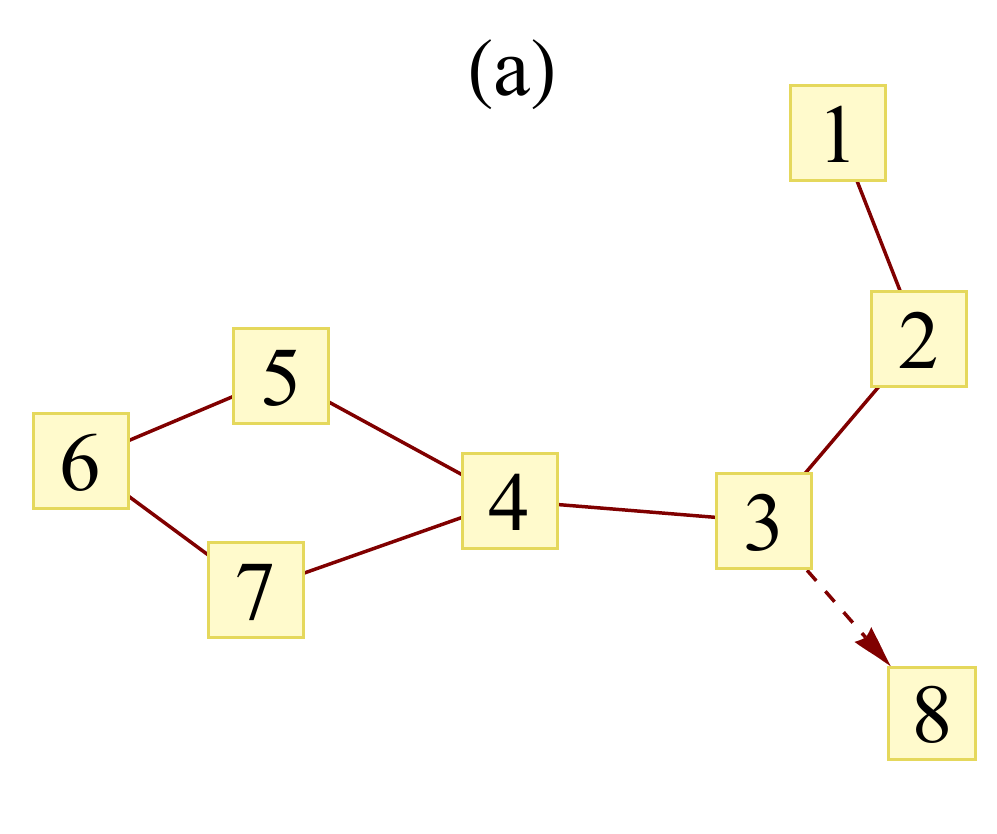}
  \includegraphics[width=0.6\textwidth,valign=c]{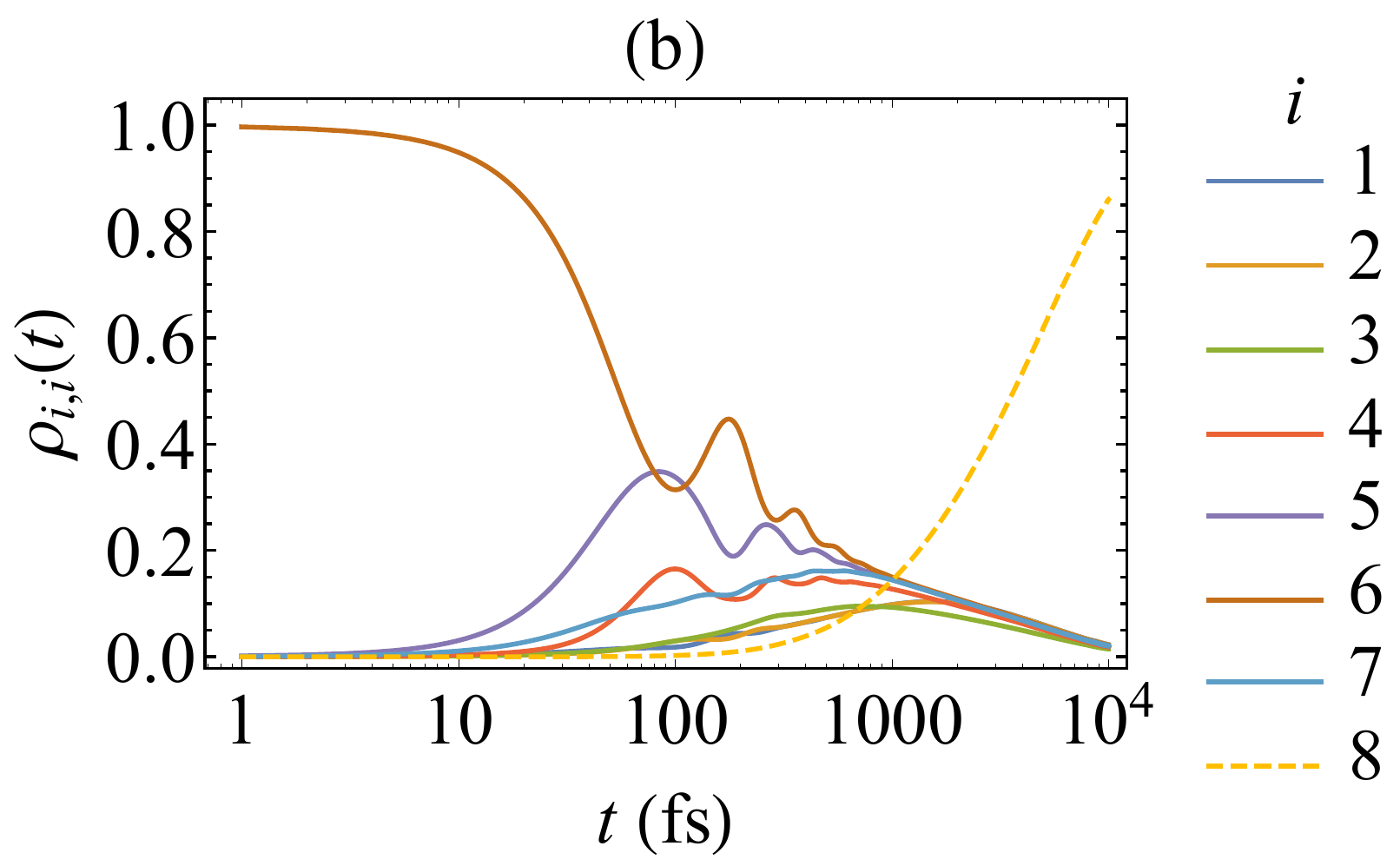}
  \caption{(a) Simplified graph structure of the FMO complex. (b) Vertex
    populations as a function of time for a QSW on the FMO model (with parameter
    values $\gamma = 1$, $\omega=0.1$, $\alpha=100$).}
\end{figure}

As a concrete example, we consider the model of Hoyer \etal \cite{hoyer2010},
which uses a Hamiltonian obtained by Adolphs and Renger
\cite{adolphs2006}:
\begin{equation} \label{eq:fmo hamiltonian}
  H = \left(\begin{array}{ccccccc}
    \mathbf{200} & \mathbf{-96} & 5 & -4.4 & 4.7 & -12.6 & -6.2 \\
    \mathbf{-96} & \mathbf{320} & \mathbf{33.1} & 6.8 & 4.5 & 7.4 & -0.3 \\
    5 & \mathbf{33.1} & \mathbf{0} & \mathbf{-51.1} & 0.8 & -8.4 & 7.6 \\
    -4.4 & 6.8 & \mathbf{-51.1} & \mathbf{110} & \mathbf{-76.6} & -14.2 & \mathbf{-67} \\
    4.7 & 4.5 & 0.8 & \mathbf{-76.6} & \mathbf{270} & \mathbf{78.3} & -0.1 \\
    -12.6 & 7.4 & -8.4 & -14.2 & \mathbf{78.3} & \mathbf{420} & \mathbf{38.3} \\
    -6.2 & -0.3 & 7.6 & \mathbf{-67} & -0.1 & \mathbf{38.3} & \mathbf{230}
  \end{array}\right).
\end{equation}
The bold entries represent the chromophores with the largest hopping
probabilities, and correspond to vertices 1 to 7 of the graph shown in
Fig.~\ref{fig:fmo}a. The units of energy are $\mathrm{cm}^{-1}$ (using the
spectroscopy convention of expressing energy in terms of the wavelength of a
photon with that energy, \ie $1\mathrm{cm}^{-1} \equiv 1.23984 \times 10^{-4}
\mathrm{eV}$). It is also customary to work in units with $\hbar=1$, so that the
unit of time is $5.309\mathrm{ps}$.

In the FMO model, the exciton is created at chromophore 6, so for our QSW we use
initial condition $\hat{\rho}(0) = \ket{6}\bra{6}$. Chromophore 3 is the site
where the exciton is absorbed into the reaction centre, which has been modeled
in previous studies by including non-probability conserving ``loss'' terms in
the master equation. Here, we take a simpler approach and add an extra vertex
(labeled 8 in Fig.~\ref{fig:fmo}a). This ``sink'' vertex is connected via a
directed edge to vertex 3, but it is not included in the Hamiltonian; it therefore
contributes incoherent scattering via a Lindblad operator but does not
participate in the coherent evolution.

Our QSW model then takes $H$ from Eq.~(\ref{eq:fmo hamiltonian}) (padded with
zeros to make an $8\times 8$ matrix) to describe the coherent evolution. For the
incoherent evolution we use the usual set of Lindblad operators from
Eq.~(\ref{eq:Lk definition}) (using $H_{ij}$ in place of $M_{ij}$), representing
dephasing scattering as well as incoherent scattering between chromophores; this
is a slight departure from Hoyer \etal, who only included the dephasing
scattering. The exciton absorption at the sink vertex is represented by an
extra Lindblad operator, $\hat{L}_{k}=\sqrt{\alpha \gamma} \ket{8} \bra{3}$,
where the factor $\alpha$ determines the rate of absorption.

In Fig.~\ref{fig:fmo}b we show the vertex populations as a function of time for
a particular set of parameters, which demonstrates excellent qualitative agreement with
Fig.~3d of Ref.~\cite{hoyer2010} (whose calculation used dephasing scattering
rates based on a temperature of 77K). A key feature is the damping of oscillations
due to the combination of coherent and incoherent dynamics. Also evident is the
accumulation of probability at the sink vertex ($i=8$), at the expense of the
remaining vertices, representing the absorption of the exciton into the reaction
centre. Although temperature does not appear explicitly in our model, its effect
is approximated by the factor $\omega$: $\omega\rightarrow 0$ is equivalent to
the low-temperature limit of a phase coherent QW, while $\omega \rightarrow 1$
corresponds to the high-temperature limit of a CRW with no phase coherence.

\subsection{Quantum Page Rank}
\label{subsection:quantum pagerank}

As an interesting application of QSWs to directed graphs, we now briefly discuss
the recent continuous-time quantum page rank algorithm (QPR) introduced by
S{\'a}nchez-Burillo \etal \cite{sanchezBurillo2012, loke2015}, based on the
well-known (classical) Page Rank algorithm (CPR) \cite{brin2012} which is a key
component of the Google search engine. The basic idea of CPR is to represent
the world wide web as a directed graph, with each webpage representing a vertex
and the links on each page representing outward edges to the pages linked to;
the algorithm then carries out a CRW on this graph to determine the centrality,
or ``rank'', of each page. In the QPR algorithm, the CRW is replaced by a QSW
which respects the directed nature of the graph.

A detailed discussion of the QPR and CPR algorithms can be found in
Refs.~\cite{sanchezBurillo2012, loke2015}; here we just describe the main
features. We begin with a directed graph $G$ having $N$ vertices and adjacency
matrix $A$. The Hamiltonian is defined as usual in terms of a symmetrized form
of $G$ (Eq.~\ref{eq:QSW hamiltonian}). The Lindblad operators are defined in
terms of the so-called ``Google matrix'', $\mathcal{G}$:
\begin{equation} \label{eq:QPR lindblad}
  \hat{L}_k = \sqrt{\gamma \mathcal{G}_{ij}} \ket{i}\bra{j},
\end{equation}
where 
\begin{equation} \label{eq:PR adjacency matrix}
  \mathcal{G}_{ij} = \begin{cases}
    \alpha A_{ij}/\mathrm{outDeg}(j) + (1-\alpha)\frac{1}{N}, &
    \mathrm{outDeg}(j) > 0, \\
    \frac{1}{N}, & \mathrm{outDeg}(j) = 0. 
    \end{cases}
\end{equation}
The effect of $\mathcal{G}$ is to avoid ``dead ends'' (nodes with out-degree 0
are scattered equally to all other vertices) and to introduce random jumps
between all vertices. The ``damping factor'' $\alpha$ controls the weighting of
the random jumps relative to the edges in the original graph ($\alpha=0.85$ is a
common choice in the literature).

As with all QSWs, the parameter $\omega$ determines the relative weighting of
coherent and incoherent components. For $\omega=0$ we get a QW on a symmetrized
graph; this does not in general provide a useful ranking algorithm, as it
does not incorporate the graph directionality. For $\omega=1$, on the other hand,
we recover the original CPR algorithm. For intermedate $\omega$, the QPR algorithm
proceeds by evaluating the QSW (Eq.~\ref{eq:QSW matrix exp}) from arbitrary
initial conditions with $t$ large enough for $\hat{\rho}(t)$ to reach a
stationary state. The vertex populations, $\rho_{ii}(t)$, then represent their
rank.

The implementation using {\tt QSWalk} is once again straightforward. The only
new step is the definition of $\hat{L}_k$ (Eq.~\ref{eq:QPR lindblad}), for which
we now use the {\tt GoogleMatrix} function:
\begin{verbatim}
  LkSet = LindbladSet[Sqrt[gamma] GoogleMatrix[G, alpha]];
\end{verbatim}
The remainder of the calculation proceeds as in
Section~\ref{subsection:transition}. It is worth noting the built-in {\it
  Mathematica} function {\tt PageRankCentrality}, which computes the CPR and is
a useful reference point.

\begin{figure}\label{fig:quantum pr}
  \includegraphics[width=0.4\textwidth,valign=c]{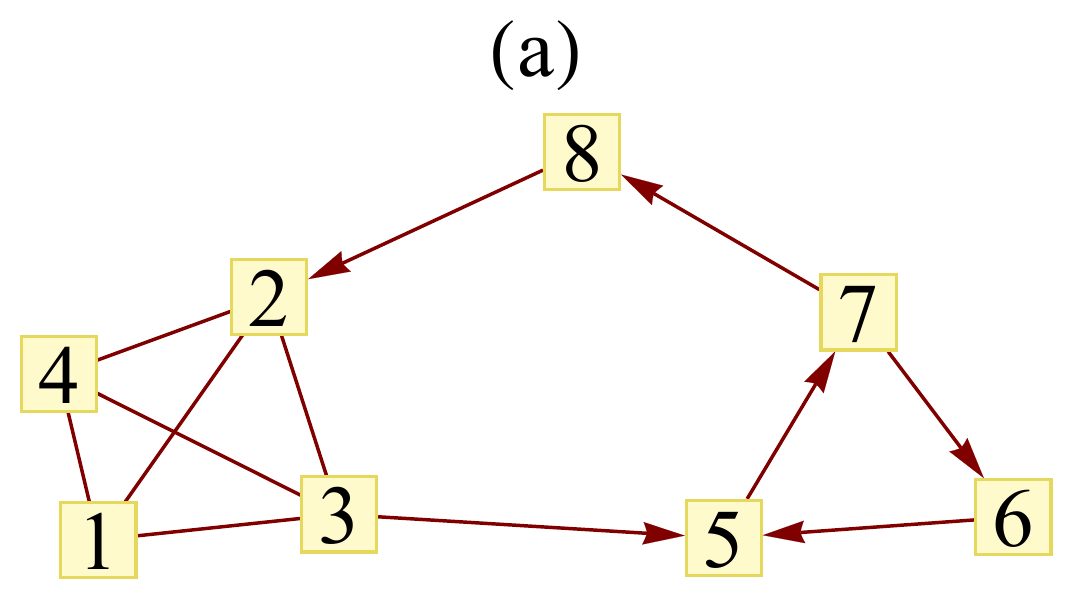}
  \hspace{0.05\textwidth}
  \includegraphics[width=0.5\textwidth,valign=c]{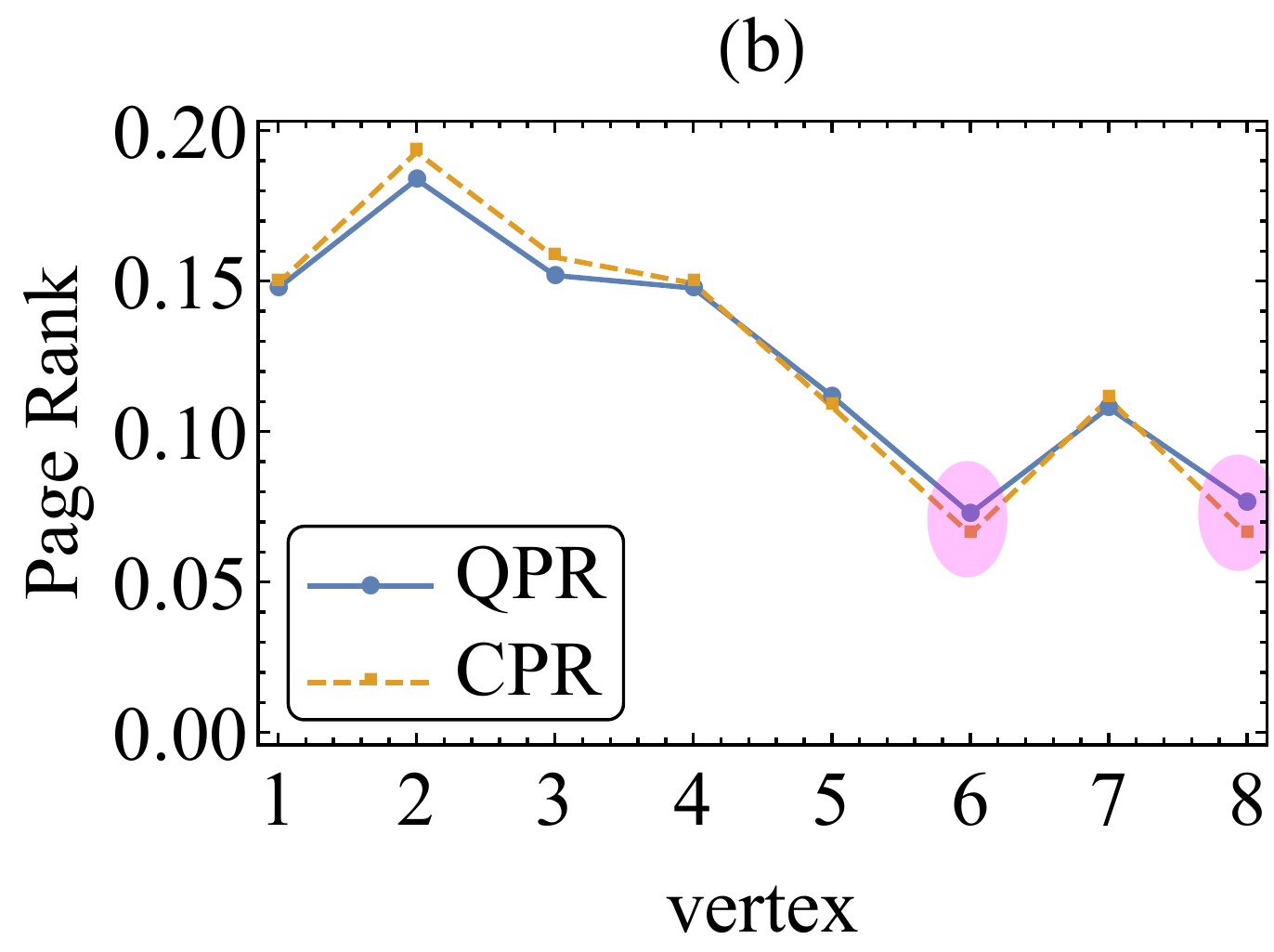}
  \caption{(a) The graph used by S{\'a}nchez-Burillo \etal
    \cite{sanchezBurillo2012} to illustrate degeneracy-breaking by the quantum
    page rank algorithm. (b) Quantum and classical Page Ranks of the vertices of
    this graph; the highlighted vertices have equal CPR but different
    QPR. (The parameters used in these calculations are $t=100$, $\gamma=1$,
    $\alpha=0.85$, $\omega=0.8$.)}
\end{figure}

As a simple illustration, in Fig.~\ref{fig:quantum pr} we show an example from
Ref.~\cite{sanchezBurillo2012}. The QPR and CPR values are broadly similar,
although the QPR calculation ``lifts'' the degeneracy of two vertices whose
ranking under CPR is equal. In Ref.~\cite{sanchezBurillo2012}, the authors show
that this effect also holds for larger real-world networks, suggesting that QPR
could become an important application for QSWs.

\section{Conclusion}

The intensive research in quantum computing and information theory over recent
decades has spurred interest in a range of quantum walk models such as the
recently-developed quantum stochastic walks. As well as the exciting potential
for practical applications down the line, these models are uncovering a wealth
of interesting physics. In practice, for all but the simplest examples much of
the insight is gained with the help of efficient numerical computing
tools. There is thus a clear need for software packages which abstract away the
computational details and allow the focus to remain on the physical application
at hand. It is with this perspective in mind that we have developed the {\tt
  QSWalk} package; we hope that it will be a useful tool, both in the hands of
researchers active in this field as well as instructors wishing to bring
cutting-edge models into a computer lab teaching environment.

On a typical modern desktop computer, the implementation provided by {\tt
  QSWalk} can be applied to graphs with a few hundred vertices with calculation
times on order of the tens of seconds, which should make it suitable for a wide
range of applications. However, there will be specialized applications involving
larger graphs where it becomes necessary to turn to high performance computing
(HPC) resources (supercomputers), which would be beyond the scope of our package
in its current form. One possibility for a future version of {\tt QSWalk},
therefore, would be to incorporate calls to a program using parallelized linear
algebra libraries running in an HPC environment; an approach along these lines
was used successfully in Ref.~\cite{izaac2015} to simulate phase-coherent quantum walks.

\section*{Acknowledgments}
The authors would like to acknowledge helpful discussions with Jingwei Tang, on the Lindblad formalism, and Tania Loke, on the Quantum Page Rank algorithm. PEF gratefully acknowledges the hospitality of the UWA School of Physics.





\section*{References}
\bibliographystyle{elsarticle-num}
\bibliography{paper}







\end{document}